\documentclass[12pt,preprint]{emulateapj}

\begin{document}
\title{Discovery of $z\sim8$ Galaxies in the HUDF from ultra-deep
  WFC3/IR Observations\altaffilmark{1}}
\author{R. J. Bouwens\altaffilmark{2,3},
  G. D. Illingworth\altaffilmark{2}, P. A. Oesch\altaffilmark{4},
  M. Stiavelli\altaffilmark{5}, P. van Dokkum\altaffilmark{6},
  M. Trenti\altaffilmark{7}, D. Magee\altaffilmark{2},
  I. Labb\'{e}\altaffilmark{8}, M. Franx\altaffilmark{3},
  C. M. Carollo\altaffilmark{4}, V. Gonzalez\altaffilmark{2} }
\altaffiltext{1}{Based on observations made with the NASA/ESA Hubble
  Space Telescope, which is operated by the Association of
  Universities for Research in Astronomy, Inc., under NASA contract
  NAS 5-26555. These observations are associated with programs
  \#11563, 9797.}  \altaffiltext{2}{UCO/Lick Observatory, University
  of California, Santa Cruz, CA 95064} \altaffiltext{3}{Leiden
  Observatory, Leiden University, NL-2300 RA Leiden, Netherlands}
\altaffiltext{4}{Institute for Astronomy, ETH Zurich, 8092 Zurich,
  Switzerland} \altaffiltext{5}{Space Telescope Science Institute,
  Baltimore, MD 21218, United States} \altaffiltext{6}{Department of
  Astronomy, Yale University, New Haven, CT 06520}
\altaffiltext{7}{University of Colorado, Center for Astrophysics and
  Space Astronomy, 389-UCB, Boulder, CO 80309, USA}
\altaffiltext{8}{Carnegie Observatories, Pasadena, CA 91101, Hubble
  Fellow}

\begin{abstract}
We utilize the newly-acquired, ultra-deep WFC3/IR observations over
the HUDF to search for star-forming galaxies at $z\sim8-8.5$, only 600
million years from recombination, using a $Y_{105}$-dropout selection.
The new 4.7 arcmin$^2$ WFC3/IR observations reach to $\sim$28.8 AB mag
($5\sigma$) in the $Y_{105}J_{125}H_{160}$ bands.  These remarkable
data reach $\sim$1.5 AB mag deeper than the previous data over the
HUDF, and now are an excellent match to the HUDF optical ACS data.
For our search criteria, we use a two-color Lyman-Break selection
technique to identify $z\sim8-8.5$ $Y_{105}$-dropouts.  We find 5
likely z$\sim$8-8.5 candidates.  The sources have $H_{160}$-band
magnitudes of $\sim$28.3 AB mag and very blue $UV$-continuum slopes,
with a median estimated $\beta$ of $\lesssim-2.5$ (where
$f_{\lambda}\propto\lambda^{\beta}$).  This suggests that $z\sim8$
galaxies are not only essentially dust free but also may have very
young ages or low metallicities.  The observed number of
$Y_{105}$-dropout candidates is smaller than the 20$\pm$6 sources
expected assuming no evolution from $z\sim6$, but is consistent with
the 5 expected extrapolating the Bouwens et al. 2008 LF results to
$z\sim8$.  These results provide evidence that the evolution in the LF
seen from $z\sim7$ to $z\sim3$ continues to $z\sim8$.  The remarkable
improvement in the sensitivity of WFC3/IR has enabled HST to cross a
threshold, revealing star-forming galaxies at z$\sim$8-9.

\end{abstract}

\section{Introduction}

An important uncharted frontier is understanding how galaxies build up
and evolve from the earliest times.  While great progress has been
made in characterizing the galaxy population at $z\lesssim6$,
extending these studies to $z\gtrsim7$ has proven extraordinarily
challenging.  Only $\sim$25 high-quality $z\sim7$ candidates are known
(e.g., Bouwens et al.\ 2008; Oesch et al.\ 2009; Ouchi et al.\ 2009;
Castellano et al.\ 2009; Gonzalez et al.\ 2009; R.J. Bouwens et
al.\ 2009, in prep).  Fundamentally, the challenge has been to obtain
extremely deep observations in the near-IR, where the redshifted $UV$
light of faint $z\gtrsim7$ galaxies is found.

Now, with the installation of the WFC3/IR camera on HST, we have a far
superior surveying instrument, with 6$\times$ the area of NICMOS,
$\gtrsim$2$\times$ the resolution, and 2-4$\times$ the sensitivity.
These capabilities allow us to search for $z\gtrsim7$ galaxies
$\sim$40$\times$ more efficiently.

Here we report on our use of the early WFC3/IR observations over the
HUDF to search for galaxies at $z\gtrsim8$.  This is the same epoch in
which a GRB was recently discovered at $z\sim8.2$ (e.g., Salvaterra et
al.\ 2009; Tanvir et al.\ 2009).  Throughout this work, we quote
results in terms of the luminosity $L_{z=3}^{*}$ Steidel et
al.\ (1999) derived at $z\sim3$, i.e., $M_{1700,AB}=-21.07$.  We refer
to the F606W, F775W, F850LP, F105W, F125W, and F160W bands on HST as
$V_{606}$, $i_{775}$, $z_{850}$, $Y_{105}$, $J_{125}$, and $H_{160}$,
respectively, for simplicity.  Where necessary, we assume $\Omega_0 =
0.3$, $\Omega_{\Lambda} = 0.7$, $H_0 = 70\,\textrm{km/s/Mpc}$.  All
magnitudes are in the AB system (Oke \& Gunn 1983).

\begin{figure}
\epsscale{1.15}
\plotone{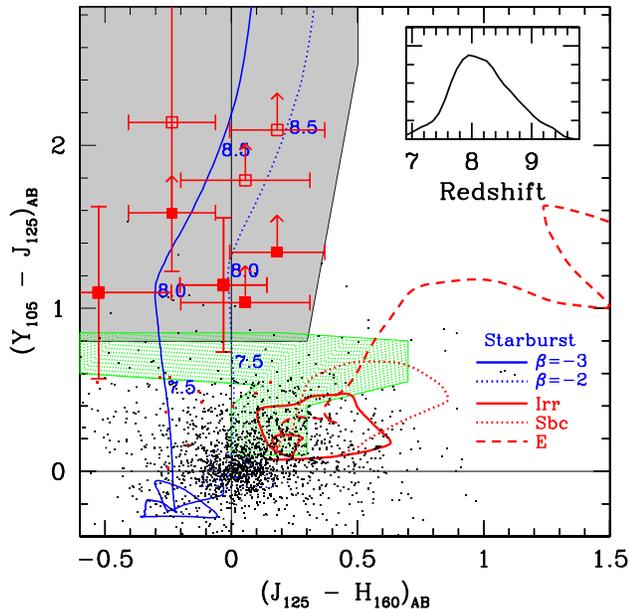}
\caption{$Y_{105}-J_{125}$ vs. $J_{125}-H_{160}$ two-color selection
  used for identifying $z\sim$8 $Y_{105}$-dropout galaxies in our HUDF
  WFC3/IR data.  The colors of our 5 $Y$-dropout candidates are
  indicated with the large red squares (with $1\sigma$ error bars and
  $2\sigma$ limits: to be conservative) and open red squares (with
  $1\sigma$ limits).  The blue lines show the colors expected for
  young star-forming galaxies (with $UV$ continuum slopes $\beta=-3$
  or $\beta=-2$) as a function of redshift.  The solid, dashed, and
  dotted red lines show the colors we would expect for various
  low-redshift galaxy SEDs (Coleman et al.\ 1980).  The hatched green
  region indicates the region in color space we would expect L,T
  dwarfs to lie (Knapp et al.\ 2004).  The colors of individual
  sources in our photometric sample are given with the small black
  points (for those detected in the HUDF optical data and hence not
  included in our selection) and the small red points (for those
  undetected in the optical).  The inset to the figure presents the
  expected redshift distribution for the sample using the simulations
  described in \S4.1 and the fiducial LFs from \S4.2.  All of our
  $z\sim8$ $Y_{105}$-dropout candidates have bluer $UV$-continuum
  slopes $\beta$ than $-2$ (as indicated by their $J_{125}-H_{160}$
  colors), with a median $\lesssim-2.5$.  This suggests that
  $z\gtrsim8$ galaxies are not only dust free but also probably have
  very young ages or low metallicities (see also Bouwens et
  al.\ 2010).\label{fig:yjjh}}
\end{figure}

\section{HUDF WFC3/IR Observations}

The present high redshift galaxy searches utilize the first epoch of
ultra-deep near-IR WFC3/IR observations acquired over the HUDF
(Beckwith et al.\ 2006) for the 192-orbit HUDF09 program (GO11563).
This program will create three ultra-deep WFC3/IR fields, one
positioned over the HUDF and the other two over the HUDF05 fields
(Oesch et al.\ 2007), each imaged in three near-IR bands $Y_{105}$,
$J_{125}$, and $H_{160}$.  Combining these ultra-deep near-IR data
with the similarly deep optical HUDF data permits us to select
$z\sim7$ $z_{850}$, $z\sim8$ $Y_{105}$, and even $z\sim10$ $J$ dropout
galaxies to very low luminosities (i.e., $-$18 AB mag, $\sim$0.06
$L_{z=3}^{*}$).

The WFC3/IR field over the HUDF is centered on 3$^h$32$^m$38.5$^s$ and
$-$27$^d$47$'$0.0$''$.  In the first year of observations, we obtained
16 orbits of $Y_{105}$-band data (2 orbits were severely impacted by
persistence and are not included), 16 orbits of $J_{125}$-band data,
and 28 orbits of $H_{160}$-band data.  The 60-orbit observations were
obtained from August 26, 2009 to September 6, 2009.

Standard techniques were used to reduce the HUDF09 WFC3/IR imaging
data.  Individual images -- after masking out sources -- were median
stacked to create super median images (one per filter) and these
median images were then subtracted from the individual frames.  The
images were then aligned and drizzled onto the same grid as the v1.0
HUDF ACS data (Beckwith et al.\ 2006) rebinned on a 0.06$''$-pixel
scale.  The drizzling was done using a modified version of
multidrizzle (Koekemoer et al. 2002).  $4\sigma$ outliers were
rejected.

Given that the WFC3/IR instrument is still relatively new, we
initially made our own estimates of the photometric zeropoints by
performing PSF-matched photometry on sources present in both the new
observations and the HUDF NICMOS observations (e.g., Thompson et
al.\ 2005; Bouwens et al.\ 2008; Oesch et al.\ 2009).  The zeropoints
derived were consistent ($<$0.05 mag) with the official STScI values,
so we elected to use the STScI values 26.27, 26.25, and 25.96 mag for
the $Y_{105}$, $J_{125}$, and $H_{160}$ bands, respectively.  The
approximate $5\sigma$ depths of the $Y$, $J$, and $H$ WFC3/IR images
are 28.8, 28.8, and 28.8 mag, respectively, in 0.4$''$-diameter
apertures, $\sim$1.5 mag deeper than the NICMOS data over the HUDF
(Thompson et al.\ 2005).  These depths were estimated by measuring the
noise statistics in apertures of various size.  The FWHM of the PSF in
our WFC3/IR near-IR images is $\sim$0.16$''$.  For reference, the HUDF
optical $B_{435}V_{606}i_{775}z_{850}$ data (Beckwith et al.\ 2006)
reached to 29.4, 29.8, 29.7, and 29.0 AB mag (5$\sigma$:
0.35$''$-diameter apertures) and had PSF FWHMs of $\sim$0.10$''$.

\begin{figure}
\epsscale{1.15}
\plotone{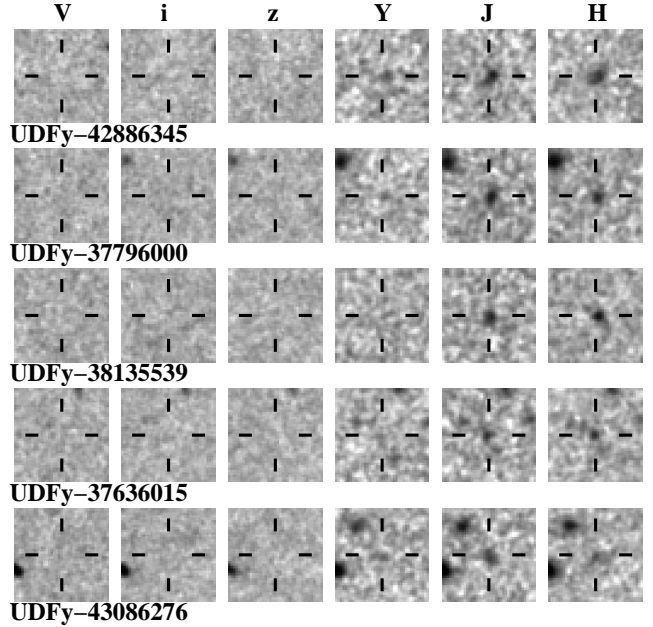}
\caption{$V_{606}i_{775}z_{850}Y_{105}J_{125}H_{160}$ image cutouts
  ($2.4''\times2.4''$ on a side) of the five $z\sim$8
  $Y_{105}$-dropout candidates we identified in the ultra-deep HUDF
  WFC3/IR observations.  All five candidates are detected at $5\sigma$
  in both the $J_{125}$ and $H_{160}$ bands, but show no detection
  ($>2\sigma$) in the ultra-deep optical HUDF $V_{606}$, $i_{775}$, or
  $z_{850}$ band images.  \label{fig:cutouts}}
\end{figure}

\begin{deluxetable*}{cccccc}
\tablewidth{11cm}
\tablecolumns{7} 
\tablecaption{$z\sim8$ $Y_{105}$-dropout candidates identified over our ultra-deep HU
DF WFC3/IR field.\label{tab:ycandlist}} 
\tablehead{ \colhead{Object ID} &
\colhead{R.A.} & \colhead{Dec} & \colhead{$H_{160}$} &
\colhead{$Y_{105}-J_{125}$\tablenotemark{a}} & \colhead{$J_{125}-H_{160}$}}
\startdata
UDFy-42886345 & 03:32:42.88 & $-$27:46:34.5 & 28.0$\pm$0.1 & 1.1$\pm$0.4 & 0.0$\pm
$0.2\\
UDFy-37796000 & 03:32:37.79 & $-$27:46:00.0 & 28.0$\pm$0.1 & 2.1$\pm$0.9 & $-$0.3$\pm
$0.2\\
UDFy-38135539 & 03:32:38.13 & $-$27:45:53.9 & 28.1$\pm$0.1 & $>$2.1 & 0.2$\pm$0.2\\
UDFy-37636015 & 03:32:37.63 & $-$27:46:01.5 & 28.4$\pm$0.1 & $>$1.8 & 0.1$\pm$0.3\\
UDFy-43086276 & 03:32:43.08 & $-$27:46:27.6 & 29.0$\pm$0.2 & 1.1$\pm$0.5 & $-$0.5$\pm
$0.3
\enddata 
\tablenotetext{a}{Lower limits on the measured colors are the $1\sigma$ limits.  Magnitudes are AB.}
\end{deluxetable*}

\section{Object Detection and Verification}

\subsection{Catalog Construction} 

Our procedure for doing object detection and photometry is identical
to that used in previous work (e.g., Bouwens et al.\ 2007; Bouwens et
al.\ 2008) and is performed using SExtractor (Bertin \& Arnouts 1996:
run in dual-image mode) on the registered data.  Object detection is
done from the coadded $J_{125}$ and $H_{160}$-band image (explicitly,
using the square root of the $\chi^2$ image: Szalay et al.\ 1999),
both of which are redward of the break for $Y_{105}$-dropout galaxies.
After smoothing the optical data to match the WFC3/IR PSFs, colors are
measured using Kron (1980)-style photometry in small apertures that
scale with the size of the source (for Kron factors of 1.2).  Flux
measurements in these small apertures are corrected up to total
magnitudes using the flux in larger scalable apertures (Kron factors
of 2.5).  This correction to total magnitudes (see e.g. Figure 5 from
Coe et al.\ 2006) is done on a source by source basis, based on the
square root of $\chi^2$ image (approximately proportional to the
coadded flux).  Finally, the total magnitudes were corrected by 0.1
mag to account for light on the wings of the PSF.

\subsection{Dropout Selection}

Sources are selected over 4.7 arcmin$^2$ with the deepest WFC3/IR
observations ($\lesssim$10\% of this area is not at the full depth).
The selection criteria we adopt for identifying $z\sim8$ $Y_{105}$
dropout galaxies are simple analogues of the criteria used to select
Lyman Break galaxies at lower redshift (e.g., Steidel et al.\ 1996;
Giavalisco et al.\ 2004; Bouwens et al.\ 2007).  That is, we require
galaxies to show strong breaks $(Y_{105}-J_{125} > 0.8)$ at the
redshifted position of Ly$\alpha$ at $z\sim8$ and to be blue redward
of the break, i.e., $(J_{125}-H_{160} < 0.5)$ and $(J_{125}-H_{160}) <
0.2 + 0.12(Y_{105}-J_{125})$ to exclude intrinsically red galaxies at
lower redshift (Figure~\ref{fig:yjjh}).  We also require our
$Y_{105}$-dropout candidates to show no detection ($<2\sigma$:
color-measurement aperture) in all bands blueward of the dropout band,
i.e., $B_{435}V_{606}i_{775}z_{850}$.  Sources showing a $>1.5\sigma$
detection in $>1$ optical band were also eliminated.  All of our
$Y_{105}$-dropout candidates were required to be $5.5\sigma$
detections in the $J_{125}$ band to ensure they corresponded to real
sources.  We elected to use a $5.5\sigma$ criterion to be conservative
for these early WFC3/IR data.

The most stringent aspect of the current selection is our requirement
that sources be undetected in the ultra-deep HUDF optical data (which
reaches to 31.5-32.0 at $1\sigma$ for most sources).  From our
simulations (\S3.4), this requirement eliminates almost all
contamination from $z\lesssim6$ galaxies, and hence we can use a
modest $(Y_{105}-J_{125})>0.8$ break to select $z\sim8$ galaxies,
without significant contamination concerns.

\begin{figure}
\epsscale{1.15}
\plotone{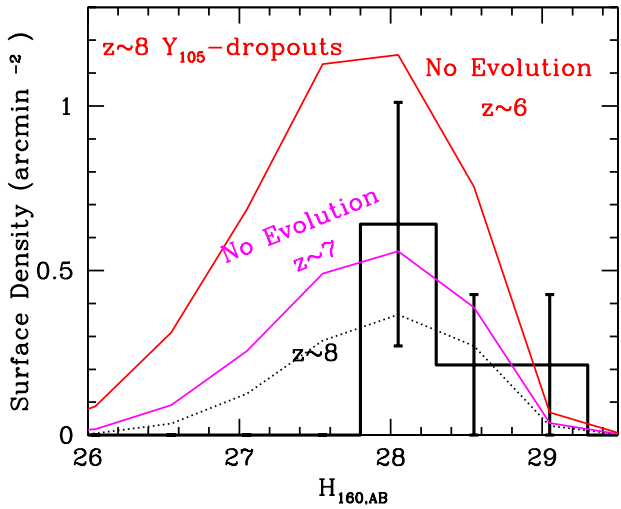}
\caption{Surface density of $z\sim8$ $Y_{105}$-dropouts found in our
  ultra-deep HUDF WFC3/IR field versus $H_{160}$-band magnitude.  For
  comparison, the surface density of $Y_{105}$-dropouts one would
  expect assuming no evolution from $z\sim6$ and $z\sim7$ is also
  plotted (\textit{shown with the red and magenta lines,
    respectively}: see \S4.1).  Also plotted (\textit{dotted black
    line}) is the expected surface density if one extrapolates the
  $z$$\sim$4-6 LF results of Bouwens et al.\ (2008) to $z\sim8$, i.e.,
  with a $\phi^*$ of $\sim$0.0011 Mpc$^{-3}$, $\alpha$ of $-1.74$, and
  a fainter value of $M^*$.
\label{fig:numc}}
\end{figure}

\begin{figure}
\epsscale{1.15}
\plotone{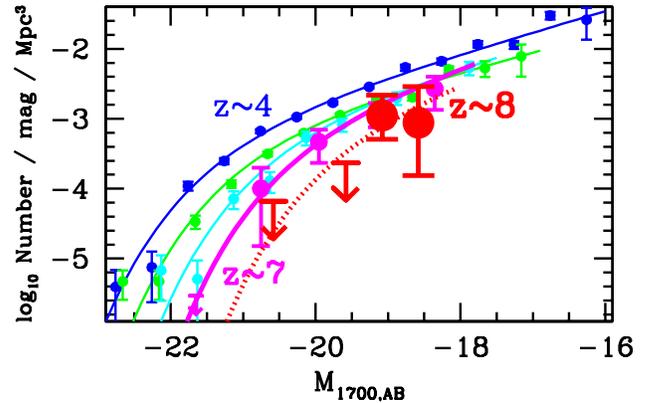}
\caption{Constraints on the rest-frame $UV$ LF at
  $z\sim8$ from the present $Y_{105}$-dropout searches over our
  ultra-deep HUDF WFC3/IR field (solid red circles, $1\sigma$ error
  bars and upper limits).  Also included is our $UV$ LF determination
  at $z\sim7$ from a $z$-dropout search over the same ultra-deep
  WFC3/IR field (magenta lines and points: Oesch et al.\ 2010a) and
  also the Bouwens et al.\ (2007) $UV$ LF determinations at $z\sim$4,5
  and 6 (shown in blue, green, and cyan lines and points).  While the
  other UV LFs shown here were derived at slightly different
  rest-frame wavelengths, their $M_{UV,AB}$ at 1700\AA$\,$is
  essentially the same ($\lesssim0.1$ mag: since star-forming dropout
  galaxies are very flat in $f_{\nu}$).  The dotted red line gives one
  possible Schechter-like LF that agrees with our search results and
  adopts the Bouwens et al.\ (2008) scaling of $M^*$ with redshift.\label{fig:lf}}
\end{figure}

\subsection{$Y_{105}$-dropout Sample}

We identify five sources that satisfy our $Y$-dropout criteria (see
Table~\ref{tab:ycandlist}).  The position of these sources in the
$Y_{105}-J_{125}$/$J_{125}-H_{160}$ two-color plane is given in
Figure~\ref{fig:yjjh}.\footnote{It is encouraging that recently both
  McLure et al.\ (2010) and Bunker et al.\ (2010) have independently
  identified a very similar set of $z\sim8$ candidates in the first
  epoch HUDF09 WFC3/IR observations over the HUDF.}  The
$J_{125}-H_{160}$ colors of our $Y_{105}$-dropout candidates are very
blue in general, corresponding to $UV$-continuum slopes $\beta$ of
$\lesssim-2$ (where $f_{\lambda}\propto\lambda^{\beta}$), with a
median $\beta\lesssim-2.5$ (though we emphasize the number of sources
are still small and the inferred $\beta$'s may be affected by
uncertainties in the photometry).  The observed $\beta$'s are blue
enough that $z\gtrsim7$ galaxies must be essentially dust free, and
possibly also have very young ages or metallicities (see also Bouwens
et al.\ 2010).

Cutouts of the sources are provided in Figure~\ref{fig:cutouts}.  The
candidates have $H_{160,AB}$ band magnitudes of $\sim$28.0-29.0 AB
mag, within one magnitude of our sensitivity limit.  Such sources
could not have been found to date since they are fainter than could be
probed with other data sets.  This illustrates the importance of the
very deep near-IR data being collected as part of this program.

All five candidates have apparent half-light radii of $\sim$0.15$''$
($\sim$0.7 $h^{-1}$kpc) -- measured using SExtractor -- not much
larger than the PSF.  The 4 candidates -- for which crowding is not a
concern -- do not show significant ($>$2$\sigma$) detections in the
IRAC data over the GOODS fields either individually or when stacked
(Labb{\'e} et al.\ 2010).  The non-detection of the candidates in the
IRAC data is not particularly surprising given the much shallower
depths of the IRAC data ($\lesssim$27.0 AB mag at $2\sigma$) relative
to the WFC3/IR data.

\subsection{Contamination Corrections}

The only meaningful source of contamination for the present sample are
sources that enter the selection via photometric scatter.  A simple
estimate of the likely contamination can be obtained by adding noise
to the color distribution observed for $\sim$25-26.5 AB mag galaxies.
The advantage of using this color distribution for the simulations is
that the distribution is realistic (being taken from the
observations), has a higher S/N than at fainter magnitudes, and does
not not include any $z\gtrsim7.5$ galaxies.  We find $\lesssim$0.2
contaminants per field, suggesting such a source of contamination is
minimal ($\lesssim$4\%).

The $<4$\% contamination estimate implicitly includes the contribution
of $z$$\sim$1.5-2 Balmer-Break Galaxies (BBG) sources scattered into
our selection.  We expect the explicit contribution to be small given the
lack of faint sources with $V-J>1.5$, $J-H>0.15$ colors expected for
BBGs (arbitrary reddenings, metallicities: R.J. Bouwens et al.\ 2010,
in prep).

Other sources of contamination are not important for this selection.
For example, given that each source is detected at $\geq$5.5$\sigma$
in the $J_{125}$ band and $\geq$4$\sigma$ in the $H_{160}$ band, no
contamination from spurious sources is expected.  Contamination by SNe
is also not important, given that the $Y_{105}J_{125}H_{160}$ data
were taken over the same 12-day window.  Finally, contamination by
stars is also unlikely.  Not only are T dwarfs rare over the
CDF-South, with a surface density $\lesssim$0.04 arcmin$^{-2}$ (e.g.,
Bouwens et al.\ 2008) and therefore unlikely to be found in our 4.7
arcmin$^2$ field, but also all 5 of our candidates appear to be
extended (having SExtractor stellarity parameters $<$0.3: see also
Oesch et al.\ 2010b) and hence not likely to be stars.

\section{Results and Implications}

\subsection{Expected Numbers}

To interpret the results of the present $z\sim8$ $Y_{105}$-dropout
selection, it is useful to estimate how many $Y_{105}$-dropouts we
might have expected if the $UV$ LF showed no evolution from $z\sim6$
and $z\sim7$.  We estimate the numbers by creating galaxy catalogs
according to the model LFs, adding artificial galaxies to the data,
and then processing the images and doing the selection in the same way
as for the real data.  For these simulations, we model the
pixel-by-pixel profiles of galaxies at $z$$\sim$8 with
similar-luminosity galaxies from the $z\sim4$ Bouwens et al.\ (2007)
HUDF $B$ dropout sample, but scaled in size to match the observed
$(1+z)^{-1}$ size-redshift scaling (Oesch et al.\ 2010b; Bouwens et
al.\ 2004, 2006; Ferguson et al.\ 2004).  Star-forming galaxies at
$z\sim7-9$ are assumed to have mean $UV$-continuum slopes $\beta$ of
$-2.5$, with a $1\sigma$ scatter of 0.4, to match the apparent colors
of $Y_{105}$-dropouts in our sample (see Figure~\ref{fig:yjjh}).

Adopting the $z\sim6$ $i$-dropout LF from Bouwens et al.\ (2007: see
also McLure et al.\ 2009) and assuming no evolution to higher
redshift, we predict 20$\pm$6 $Y_{105}$-dropouts in our WFC3/IR field.
The $z\sim6$ predictions are shown as a function of $H_{160}$-band
magnitude in Figure~\ref{fig:numc} and compared with the observations.
We expect $\sim$70\% uncertainties in these numbers due to small
number statistics and large-scale structure variations (assuming a 4.7
arcmin$^2$ survey field, $\Delta z \sim 0.8$ redshift selection
window, and $\sim10^{-3}$ Mpc$^{-3}$ source density: e.g., Trenti \&
Stiavelli 2008).  The observed numbers are $>2\sigma$ lower than
expected from the well-determined $z\sim6$ LF.  Predicted surface
densities can also be made using the Bouwens et al.\ (2008) $z\sim7$
LF (also shown), but the uncertainties are large given the small
sample in that paper (and the new WFC3/IR observations permit much
better $z\sim7$ LF estimates).

\begin{deluxetable}{lcccc}
\tablewidth{0pt}
\tabletypesize{\footnotesize}
\tablecaption{$UV$ Luminosity Densities and Star Formation Rate
  Densities.\tablenotemark{a}\label{tab:sfrdens}}
\tablehead{
\colhead{} & \colhead{} & \colhead{$\textrm{log}_{10} \mathcal{L}$} & \multicolumn{2}
{c}{$\textrm{log}_{10}$ SFR density} \\
\colhead{Dropout} & \colhead{} & \colhead{(ergs s$^{-1}$} & \multicolumn{2}{c}{($M_{\
odot}$ Mpc$^{-3}$ yr$^{-1}$)} \\
\colhead{Sample} & \colhead{$<z>$} & \colhead{Hz$^{-1}$ Mpc$^{-3}$)} & \colhead{Uncorrected} 
& \colhead{Corrected\tablenotemark{b}}}
\startdata
$z$ & 6.8 & 25.73$\pm$0.16\tablenotemark{c} & $-2.17\pm0.16$ & $-2.17\pm0.16$ \\
$Y$ & 8.2 & 25.18$\pm$0.24 & $-2.72\pm0.24$ & $-2.72\pm0.24$
\enddata
\tablenotetext{a}{Integrated down to $-18.3$ AB mag, or 0.08 $L_{z=3}^{*}$.}
\tablenotetext{b}{The dust correction is taken to be 0, given the
very blue $\beta$'s (see also Bouwens et al. 2009; Bouwens et
al.\ 2010)}
\tablenotetext{c}{Using the $z\sim7$ $UV$ LF derived from the same
   ultra-deep WFC3/IR field as this study (Oesch et al.\ 2010a)}
\end{deluxetable}

\begin{figure}
\epsscale{1.15}
\plotone{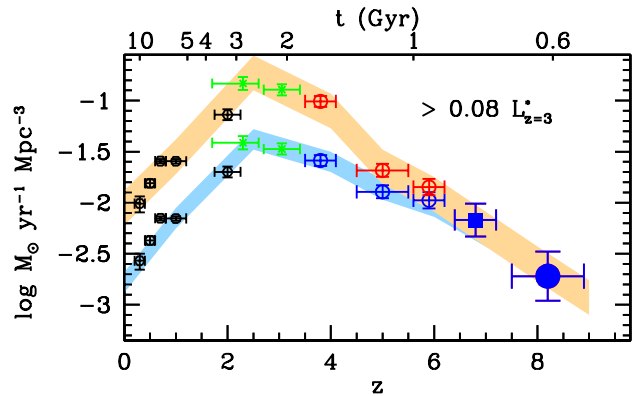}
\caption{Determinations of the $UV$ luminosity density and SFR
  density, integrated to 0.08 $L_{z=3}^{*}$ ($-$18.3 AB mag) as
  appropriate for the $z\sim8$ sample, as a function of redshift.  The
  large red circle shows the constraints we can set on this density at
  $z\sim8$ from the current $Y_{105}$-dropout search (see also
  \S4.3).  The lower set of points (and blue region) show the SFR
  density determintion inferred directly from the $UV$ light, and the
  upper set of points (and orange region) show what one would infer
  using dust corrections inferred from the $UV$-continuum slope
  measurements (e.g., Bouwens et al.\ 2009; Bouwens et al.\ 2010).
  Note that the dust correction is essentially zero at $z>6$.  Also
  included on this figure are the determinations at $z\sim7$ from the
  HUDF WFC3/IR $z$-dropout search (Oesch el.\ 2010a: \textit{solid red
    square}), the Bouwens et al.\ (2007) determination at $z\sim4-6$
  (\textit{open red circles}), the Reddy \& Steidel (2009) determinations
  at $z\sim$2-3 (\textit{green crosses}), and the Schiminovich et
  al.\ (2005) determinations at $z\lesssim2$ (\textit{black
    hexagons}).  A systematic increase in the SFR density from
  $z\sim8$ to $z\sim2$ is clear.\label{fig:sfz}}
\end{figure}

\subsection{Implications for the $z\sim8$ $UV$ LF}

The fact that the observed numbers are lower than predicted assuming
no-evolution from $z\sim6$ suggests that the UV LF continues to evolve
from $z\sim8$.  So, it is interesting to extrapolate the LF results
from Bouwens et al.\ (2008) to $z\sim8$ -- giving $M_{UV}^* = -19.45$,
$\phi^*\sim0.0011$ Mpc$^{-3}$, and $\alpha\sim-1.74$ -- and see what
we find.  Performing this exercise, we predict that 5
$Y_{105}$-dropouts would be found in the present search (\textit{shown
  with the dotted black line in the lower panel of
  Figure}~\ref{fig:lf}).  This is in good agreement with the observed
results.

We can further quantify the overall magnitude of this evolution.
Using the Bouwens et al.\ (2008) LF parameterization as a guide, we
fix $\alpha=-1.74$, $\phi^* = 0.0011$ Mpc$^{-3}$, and then derive
confidence intervals on $M_{UV}^{*}$.  For our $Y_{105}$-dropout
search, we estimate that $M_{UV}^{*} = -19.5\pm0.3$ AB mag.  This is
significantly fainter than the $M_{UV}^{*} = -20.2\pm0.2$ AB mag
estimated at $z\sim6$ or the $M_{UV}^{*} = -21$ AB mag at $z\sim4$.

Of course, given the size of the sample and lack of complementary
wide-area searches for bright $z\sim8$ sources, it is difficult to
constrain the shape of the $z\sim8$ $UV$ LF.  Therefore, we simply
consider the stepwise LF at $z\sim8$.  We divide our dropout sample
into 0.5 mag bins, compute the equivalent absolute magnitudes in each
of these bins, and then divide the observed number of sources in each
bin by the effective selection volume, which are estimated using the
same simulations described in \S4.1.  These stepwise LF
determinations are presented in Figure~\ref{fig:lf}, with the LFs at
$z$$\sim$4-7 (Bouwens et al.\ 2007; Oesch et al.\ 2010a) shown for
context.  The $1\sigma$ upper limits on the volume density of luminous
$z\sim8$ sources are also shown.  It would appear that the $UV$ LF
only shows very weak evolution at low luminosities ($\sim-18.3$ AB
mag).  This is in contrast to the dramatic evolution observed at the
bright end from $z\sim7$ to $z\sim4$ (see e.g. discussion in Shimasaku
et al.\ 2005; Bouwens et al.\ 2008).

\subsection{Constraints on the $UV$ Luminosity Density/SFR density at
  $z\sim8$}

Finally, we calculate the luminosity densities (and unobscured SFR
densities) at $z\sim8$ implied by these constraints on the rest-frame
UV LF.  For the luminosity density at $z\sim8$, we simply integrate
the stepwise $z\sim8$ LF shown in Figure~\ref{fig:lf}.  We convert
these $UV$ luminosity densities into the equivalent unobscured SFR
densities using the Madau et al.\ (1998) prescription.  The results
are presented in Figure~\ref{fig:sfz} and Table~\ref{tab:sfrdens}.
Also presented are the star formation rate densities inferred.  The
dust correction is taken to be 0, given the very blue $\beta$'s (see
also Bouwens et al. 2009; Bouwens et al.\ 2010).
\vspace{0.2cm}

As these results demonstrate, the remarkable improvement in the
sensitivity and the ``discovery efficiency'' (area gain $\times$
sensitivity gain) of WFC3/IR has enabled HST to cross a threshhold.
HST can now find star-forming galaxies at z$\sim$8-8.5.  The existence
of such galaxies and the active star formation implied at even higher
redshifts $z>10$ provides a striking framework for future detailed
JWST observations.

\acknowledgements

We acknowledge our program coordinator William Janusweski for his
exceptional care in helping to set up our program and observing
configuration.  We are grateful to all those at NASA, STScI and
throughout the community who have worked so diligently to make Hubble
the remarkable observatory that it is today. The servicing missions,
like the recent SM4, have rejuvenated HST and made it an
extraordinarily productive scientiﬁc facility time and time again.  We
greatly appreciate the support of policymakers, and all those in the
flight and servicing programs who contributed to the repeated
successes of the HST servicing missions.  We acknowledge the support
of NASA grant NAG5-7697 and NASA grant HST-GO-11563.01.  PO
acknowledges support from the Swiss National Foundation.

\end{document}